\documentclass[aps,pra,amsmath,amssymb,floatfix,footinbib,superscriptaddress,letterpaper,reprint]{revtex4-1}
\usepackage{amsmath,amssymb}
\usepackage{subfigure,graphicx,psfrag,transparent,epstopdf}
\usepackage{bm,color,anyfontsize}
\usepackage{natbib,hyperref}
\usepackage{cleveref}

\hypersetup{colorlinks=true,linkcolor=blue,citecolor=blue,urlcolor=blue}

\newcommand{\be}{\begin{equation}}
\newcommand{\ee}{\end{equation}}
\newcommand{\nn}{\nonumber}

\newcommand{\at}{\tilde{a}}
\newcommand{\alpt}{\tilde{\alpha}}

\newcommand{\dg}{^{\dagger}}
\newcommand{\Dw}{\delta \omega}
\newcommand{\epe}{\epsilon_{\mathrm{eff}}}
\newcommand{\ka}{\kappa}

\newcommand{\s}{\sigma}
\newcommand{\Tr}{\mathrm{Tr}}
\newcommand{\ze}{\zeta}
\newcommand{\lf}{\left}
\newcommand{\rt}{\right}
\newcommand{\rem}{f_j}
\newcommand{\Ran}{\mathfrak{R}}
\newcommand{\Ker}{\mathfrak{N}}

\newcommand{\abs}[1]{\left| #1 \right|}
\newcommand{\avg}[1]{\left< #1 \right>}
\newcommand{\savg}[1]{\langle #1 \rangle}
\newcommand{\nInverse}[1]{#1^{-1}}
\newcommand{\pinv}{^{\leftharpoonup 1}}
\newcommand{\sNb}[1]{\sN\lf[ #1 \rt]}
\newcommand{\Trb}[1]{\Tr \left[ #1 \right]}
\newcommand{\Range}[1]{\mathrm{Ran} \left(#1 \right)}

\newcommand{\sD}{\mathbb{D}}
\newcommand{\sL}{\mathbb{L}}
\newcommand{\sN}{\mathbb{N}}
\newcommand{\sP}{\mathbb{P}}
\newcommand{\sZ}{\mathbb{Z}_0}

\newcommand{\inner}[2]{\left<#1,#2\right>}
\newcommand{\sinner}[2]{\langle #1,#2 \rangle}

\newcommand{\De}[1]{\Delta_{#1}}
\newcommand{\Dem}{\De{\mu}}
\newcommand{\Demc}[1]{\Dem^{\lf( #1\rt)}}

\newcommand{\eva}[1]{\lambda_{#1}}
\newcommand{\evl}[1]{w_{#1}}
\newcommand{\evr}[1]{u_{#1}}

\newcommand{\evac}[2]{\eva{#1}^{\lf(#2 \rt)}}
\newcommand{\evlc}[2]{\evl{#1}^{\lf(#2 \rt)}}

\newcommand{\evam}{\eva{\mu}}
\newcommand{\evlm}{\evl{\mu}}
\newcommand{\evrm}{\evr{\mu}}

\newcommand{\evamc}[1]{\evam^{\lf(#1 \rt)}}
\newcommand{\evlmc}[1]{\evlm^{\lf(#1 \rt)}}
\newcommand{\evrmc}[1]{\evrm^{\lf(#1 \rt)}}

\newcommand{\rs}{\rho_{\mathrm{s}}}
\newcommand{\rsc}[1]{\rs^{\lf( #1\rt)}}

\newcommand{\rsa}[1]{\rs^{\mathrm{a};#1}}
\newcommand{\rsaAM}[1]{\rho_{\mathrm{s},\mathrm{AM}}^{\mathrm{a};#1}}

\newcommand{\zs}{\ze_{\mathrm{s}}}
\newcommand{\zsc}[1]{\zs^{\lf(#1\rt)}}
\newcommand{\zsa}[1]{\ze_{\mathrm{s}}^{\mathrm{a};#1}}

\newcommand{\ket}[1]{\left| #1 \right>} 
\newcommand{\bra}[1]{\left< #1 \right|}
\newcommand{\braket}[2]{\left< #1 \vphantom{#2} \right| \left. #2 \vphantom{#1} \right>} 
\newcommand{\sket}[1]{| #1 \rangle} 
\newcommand{\sbra}[1]{\langle #1 |}
\newcommand{\sbraket}[2]{\langle  #1 \vphantom{#2} | #2 \vphantom{#1} \rangle} 

\newcommand{\ps}[1]{\psi_{#1}}
\newcommand{\psm}{\ps{\mu}}
\newcommand{\psmc}[1]{\psi_{\mu}^{\lf( #1 \rt)}}

\newcommand{\ep}[1]{E_{#1}}
\newcommand{\epm}{\ep{\mu}}
\newcommand{\epmc}[1]{\epm^{\lf( #1\rt)}}
\newcommand{\demc}[1]{\delta_{\mu}^{\lf( #1\rt)}}

\definecolor{dark_green}{RGB}{0,140,0}
\begin{document}
\title{Perturbative approach to Markovian open quantum systems}
\author{Andy C.Y.~Li}
\affiliation{Department of Physics and Astronomy, Northwestern University, Evanston, Illinois 60208, USA}
\author{F. Petruccione}
\affiliation{Quantum Research Group, School of Chemistry and Physics, University of KwaZulu-Natal, Durban 4001, South Africa and National Institute for Theoretical Physics (NITheP), KwaZulu-Natal, South Africa}
\author{Jens Koch}
\affiliation{Department of Physics and Astronomy, Northwestern University, Evanston, Illinois 60208, USA}
\date{\today}

\begin{abstract}
Perturbation theory (PT) is a powerful and commonly used tool in the investigation of closed quantum systems. In the context of open quantum systems, PT based on the Markovian quantum master equation is much less developed. The investigation of open systems mostly relies on exact diagonalization of the Liouville superoperator or quantum trajectories. In this approach, the system size is rather limited by current computational capabilities. Analogous to closed-system PT, we develop a PT suitable for open quantum systems. This proposed method is useful in the analytical understanding of open systems as well as in the numerical calculation of system properties, which would otherwise be impractical.
\end{abstract}

\maketitle

\section{Introduction}
In many fields of physics, there is a great interest in understanding open quantum systems, ranging from problems in atomic and molecular physics \cite{AMO1,AMO3,AMO2} to more recent application in circuit QED \cite{circuitQED3,circuitQED4,circuitQED1,circuitQED2} or optomechanics \cite{optomechanics3,optomechanics1,optomechanics2}. The theoretical description of open quantum systems tends to be much more challenging than that of closed quantum systems. Simulations of open systems are more demanding and analytical and numerical approaches are less developed in the open-system case than for the closed-system case. This makes theoretical studies of medium and large open systems, e.g.\ open-system quantum simulators \cite{quantum_simulator_1} and dissipative phase transitions \cite{dissipative_phase_1}, particularly difficult.

Open quantum systems can often be treated by the Markovian quantum master equation \cite{breuer2002},
\be
\label{eq:master_equation}
\dot{\rho}(t)=\sL \rho(t),
\ee
which involves the density matrix $\rho$ and the generalized Liouville superoperator $\sL$. \Cref{eq:master_equation} affords the investigation of steady-state physics as well as system dynamics. One may investigate the steady state and the dynamics by exact diagonalization of $\sL$ or by simulating quantum trajectories \cite{quantum_trajectories1}. However, the system size is rather limited given current computational capabilities. Other approaches include the matrix product method \cite{MPS_1,MPS_2,MPO_1}. Although it is suitable for larger systems, the method is easily applicable only to one-dimensional systems.

For closed systems, perturbative treatments are often employed to obtain approximate results of large systems. One of the perturbative treatments is known as adiabatic elimination or Schrieffer-Wolff formalism, which can be generalized to open systems to obtain an effective Liouville superoperator \cite{SW_1,SW_2,SW_3}. This method is mostly applied to Liouville superoperators in which the spectrum can be easily separated into slow and fast subspaces. Alternatively, we consider PT that directly determines the perturbative corrections to the eigenstates and eigenvalues. PT for open systems was previously studied \cite{perturbation2011,perturbation2013}. However, results in Refs.\ \cite{perturbation2011,perturbation2013} were limited to the steady state. Also, non-positivity of steady-state results from PT due to truncation was not addressed by the authors. Here, we develop a density-matrix PT that is applicable to all eigenstates of $\sL$ including the steady state. We further construct a PT based on the amplitude matrix which yields a positive steady-state density matrix.

This paper is organized as follows. In \cref{sec:generalformalism}, we discuss the density-matrix PT and the non-positivity issue of the steady-state result due to truncation. In \cref{sec:amplitude_matrix}, we present the amplitude-matrix PT which reconstructs a positive density matrix from the density-matrix PT. In \cref{sec:examples}, we apply second-order PT to two examples to illustrate the use and accuracy of the PT.

\section{Density-matrix PT}
\label{sec:generalformalism}
In this section, we propose a non-degenerate density-matrix PT based on the quantum master equation shown in \cref{eq:master_equation}. The Liouville superoperator $\sL$ which serves as the generator of a quantum dynamical semigroup \cite{breuer2002} is not Hermitian, i.e.\ the adjoint superoperator $\sL\dg$ is not equal to $\sL$. The right and left eigenstates $\evrm$ and $\evlm$ of $\sL$ are defined by
\be
\sL \evrm=\evam \evrm
\label{eq:eigensystem_L}
\ee
and $\sL\dg \evlm=\lf(\evam\rt)^* \evlm$, respectively. Here, $\evam$ is the corresponding eigenvalue which is in general complex and $\mu$ is a non-negative integer labeling the eigenstates. With appropriate normalization, the left and right eigenstates obey the following bi-orthonormal relation:
$\inner{\evl{\mu}}{\evr{\nu}} = \delta_{\mu \nu}$.
Here, $\delta_{\mu\nu}$ is the Kronecker delta and $\inner{x}{y} \equiv \Trb{x\dg y}$ is the Hilbert-Schmidt inner product \cite{inner_product_1996}. The right eigenstates $\evrm$ together with $\evam$ consist of the information of the steady state (labeled by $\mu=0$) and the dynamics of the system.

The density-matrix PT is developed based on the series expansions of $\evam$ and $\evrm$ analogous to the case of closed-system PT:
\be
\evam = \sum_{j=0}^{\infty} \alpha^j \evamc{j} \text{, \ }
\evrm = \sum_{j=0}^{\infty} \alpha^j \evrmc{j}.
\label{eq:ev_expansion}
\ee
Here, $\evamc{j}$ and $\evrmc{j}$ are the $j$-th-order terms of the eigenvalue and the right eigenstate and $\alpha$ is a dimensionless parameter introduced for order counting. To construct the density-matrix PT, we separate $\sL$ into two parts: the unperturbed superoperator $\sL_0$ and the perturbation $\alpha \, \sL_1$, i.e.\ 
\be
\sL=\sL_0+\alpha \, \sL_1.
\label{eq:sLdecomp}
\ee
We separate $\sL$ in such a way that $\sL_0$ is a proper generator of the quantum dynamical semigroup. In addition, we choose $\sL_0$ to be solvable meaning that the part of the spectrum $\{ \evamc{0} \}$ that we are interested in and the corresponding left and right eigenstates $\{\evlmc{0}\}$ and $\{\evrmc{0}\}$ of $\sL_0$ are known. We assume that this part of the spectrum is non-degenerate. We determine recursive relations for $\evamc{j}$ and $\evrmc{j}$ by plugging \cref{eq:ev_expansion,eq:sLdecomp} into \cref{eq:eigensystem_L} and examining the result order by order in $ \alpha$. For $j$-th order in $\alpha$, we get
\be
\label{eq:linear_algebra_recursive}
\left(\sL_0 - \evamc{0} \right) \evrmc{j}  = -\sL_1 \evrmc{j-1} +\Demc{j},
\ee
where $\Demc{j}=\sum_{k=1}^{j}\evamc{k} \evrmc{j-k}$.

Up until now, the treatment is very similar to the procedure in deriving the well-known form of stationary PT for a closed system. Specifically, consider replacing $\sL_0$, $\sL_1$, $\evrm$ and $\evam$ by the unperturbed Hamiltonian $H_0$, the perturbation $V$, the eigenvectors $\sket{\psm}$ and eigenenergies $\epm$ of $H\equiv H_0+\alpha V$ respectively. Then, we find that \cref{eq:linear_algebra_recursive} has exactly the same form as the usual recursive equation in closed-system PT, i.e.\
\be
\lf( H_0 - \epmc{0}  \rt) \ket{\psmc{j}}  = -V \ket{\psmc{j-1}} +\demc{j},
\label{eq:cs_recursive}
\ee
where $\demc{j}=\sum_{k=1}^{j}\epmc{k} \sket{\psmc{j-k}}$. To obtain the recursive relation for the correction $\epmc{j}$ to the eigenenergy, we multiply \cref{eq:cs_recursive} with $\sbra{\psmc{0}}$ from the left. This yields
\be
\epmc{j} = \bra{\psmc{0}} V \ket{\psmc{j-1}} -\sum_{k=1}^{j-1}\epmc{k}  \braket{\psmc{0}}{\psmc{j-k}}.
\label{eq:eigenenergy_j}
\ee
Analogously, we take the inner product on \cref{eq:linear_algebra_recursive} with the left eigenstate $\evlmc{0}$, we obtain the recursive relation of $\evamc{j}$, 
\be
\label{eq:evaj}
\evamc{j} =  
\inner{\evlmc{0}}{\sL_1 \evrmc{j-1}} -\sum_{k=1}^{j-1}\evamc{k} \inner{\evlmc{0}}{\evrmc{j-k}}.
\ee
Note that \cref{eq:eigenenergy_j} is often further simplified by requiring $\sbraket{\psmc{0}}{\psmc{j}} =0$ for $j\neq0$. Here, we keep the corresponding term $\sinner{\evlmc{0}}{\evrmc{j-k}}$ in \cref{eq:evaj} for which the reason will be clear soon.

We next turn to the computation of the eigenstate corrections. The Hamiltonian of any closed system is Hermitian and hence $H_0$ provides a complete eigenbasis. As a result, $\sket{\psmc{j}}$ in \cref{eq:cs_recursive} can be expanded in the eigenbasis $\{ \sket{\psmc{0}}\}$ of $H_0$. Solving \cref{eq:cs_recursive} is then straightforward. However, $\sL$ is \textit{not} Hermitian and it may \textit{not} even be diagonalizable. As a result, the expansion of $\evrmc{j}$ in terms of $\evrmc{0}$ may in general fail. We therefore adopt the different strategy of finding an ``inverse" (generalized inverse) of $(\sL_0 - \evamc{0})$. Since $(\sL_0 - \evamc{0})$ is singular, we employ the Moore-Penrose pseudoinverse (which for a given matrix $A$ we denote by $A\pinv$). This pseudoinverse resembles the normal inverse but is well-defined even for non-invertible matrices. From this, we obtain
\begin{align}
\label{eq:evrj}
\evrmc{j} = \left(  \sL_0 -\evamc{0} \right)\pinv \left(-\sL_1 \evrmc{j-1}+\Demc{j} \right).
\end{align}
A review of the Moore-Penrose pseudoinverse and details of the derivation of \cref{eq:evrj} are provided in \cref{app:proofperturbation,app:pseudoinverse}. We emphasize that this pseudoinverse does not guarantee that $\sinner{\evlmc{0}}{\evrmc{j}}=0$.

The steady-state density matrix $\rs \equiv \evr{0}$ defined by $\sL \rs=0$ is of particular interest. As a special case of \cref{eq:evaj,eq:evrj}, we can simplify the corrections $\evac{0}{j}$ and $\rsc{j}$ to
\begin{align}
\label{eq:ev0j}
\evac{0}{j} &= 0, \\
\label{eq:rsj}
\rsc{j} &= -\sL_0\pinv  \sL_1 \rsc{j-1}.
\end{align}
Details of the simplification are shown in \cref{app:proofperturbation}. Corrections to $\eva{0}$ and $\rs$ were previously derived \cite{perturbation2011} without using the Moore-Penrose pseudoinverse. The result for the density-matrix corrections in Ref.\ \cite{perturbation2011} differs from ours [\cref{eq:rsj}] merely by a shift,
\be
\label{eq:shift_rsj}
\rsc{j} \rightarrow \rsc{j} + c_j \rsc{0},
\ee
where $c_j$ is a constant.
Note that the solution of \cref{eq:linear_algebra_recursive} is not unique since $(\sL_0 - \evamc{0})$ is non-invertible. We infer from $\sL_0 \rsc{0}=0$ that the shifted $\rsc{j}$ in \cref{eq:shift_rsj} is also the solution of \cref{eq:linear_algebra_recursive}. By using one particular generalized inverse, we select one solution from infinitely many solutions. The difference between the result in Ref.\ \cite{perturbation2011} and ours merely reflects the different choices of generalized inverses. Nonetheless, the two results are equivalent since shifts of the form in \cref{eq:shift_rsj} only affect the overall normalization of $\rs$ \footnote{If truncation of the series of $\rs$ is present as discussed later, the two results are different by higher-order terms. This difference, however, is insignificant since it is a higher-order effect.}. This difference in the overall normalization does not affect the result because we are going to normalize $\rs$ as follows.
Note that the steady-state density matrix obtained by summation of $\rsc{j}$ [\cref{eq:rsj}] is not normalized. As usual, we can normalize the result manually by evaluating
\be
\rs =\sNb{\textstyle\sum_{j=0}^{\infty} \alpha^j \rsc{j}}.
\label{eq:normalizedperturbedss}
\ee
Here, $\sNb{A} \equiv A/\Trb{A}$ denotes a normalization operation that normalize any non-traceless matrix $A$ to have a trace norm one.

In most situations of interest, the series in \cref{eq:normalizedperturbedss} is truncated to finite order just as in closed-system PT. Let us denote the approximate result up to $M$-th order as $\rsa{M} \equiv \sNb{\sum\nolimits_{j=0}^{M} \alpha^j \rsc{j}}$. To check for consistency, we assess whether $\rsa{M}$ indeed represents a proper density matrix which must be normalized, Hermitian and positive-semidefinite \cite{breuer2002}. By virtue of $\sN$, $\rsa{M}$ is explicitly normalized. Hermiticity can be verified by noticing that $\sL_0\pinv$ and $\sL_1$ map Hermitian operators to Hermitian operators. Due to the omission of higher-order terms in the truncation, however, positivity of $\rsa{M}$ is not guaranteed. In the examples presented in \cref{sec:examples}, this issue indeed occurs for certain parameter choices. This makes a key difference between closed-system PT and density-matrix PT. In closed-system PT, the approximate result is always a proper quantum state. However in the density-matrix PT, the approximate result may not be a proper density matrix.

Similar issues with approximations of the density matrix which violate positivity are encountered in quantum tomography due to measurement errors. There, a maximum-likelihood method is used to reconstruct a physical density matrix from the non-positive approximation \cite{maximum-likelihood_reconstruction1, maximum-likelihood_reconstruction2}. However, this method is difficult to apply to the large density matrices we are interested in. We will discuss an alternative method which suits large density matrices in next section.

\section{Amplitude-matrix PT}
\label{sec:amplitude_matrix}
In this section, we propose an amplitude-matrix PT to reconstruct a positive steady-state density matrix. Any density matrix $\rho$, which is Hermitian and positive-semidefinite, can be decomposed in the form: $\rho= \ze \ze\dg$ \cite{matrix_analysis_1990}. Here, we call $\ze$ the amplitude matrix following Ref.\ \cite{amplitude_matrix2013}. The above decomposition is not unique, meaning that there are many choices for $\ze$ which lead to the same density matrix $\rho$. To eliminate these extra degrees of freedom, we choose $\ze$ to be lower triangular with real and non-negative diagonal elements. The existence and uniqueness of $\ze$ are then guaranteed by the Cholesky decomposition \footnote{Strictly speaking, if one of the eigenvalues of $\rho$ is exactly zero, the Cholesky decomposition is non-unique and numerically unstable. We can bypass this issue by using the correction matrix mentioned in \cref{app:series_ze}.}.

Let us assume that the steady-state amplitude matrix $\zs$ can be written as a power series in $\alpha$: $\zs = \sum_{j=0}^{\infty} \alpha^j \zsc{j}$. Here, all matrices $\zsc{j}$ are again of lower triangular shape. By collecting terms of the same order in $\alpha$ from $\rs=\zs \zs\dg$, we obtain
\begin{align}
\label{eq:zs_0}
\zsc{0} \lf(\zsc{0}\rt)\dg=& \rsc{0}, \\
\label{eq:zs_j}
\sZ \zsc{j} =& \rsc{j}- \sum_{k=1}^{j-1} \zsc{k}  \lf(\zsc{j-k}\rt)\dg,
\end{align}
where $\sZ$ is is defined by $\sZ \bullet = \zsc{0} (\bullet)\dg + \bullet (\zsc{0} )\dg $. Here, $\zsc{0}$ is obtained from \cref{eq:zs_0} by Cholesky decomposition and $\zsc{j}$ is determined from the system of linear equations in \cref{eq:zs_j}. We determine $\rsc{j}$ in \cref{eq:zs_j} by density-matrix PT and thus the amplitude-matrix PT is based on the density-matrix PT.

Once again, we truncate the amplitude matrix to $M$-th order: $\zsa{M} \equiv \sum_{j=0}^{M} \alpha^j \zsc{j}$. We can then determine the steady-state density matrix $\rsaAM{M}$ by
\be
\rsaAM{M} = \sNb{\zsa{M} \lf(\zsa{M}\rt)\dg}.
\ee
Here, $\rsaAM{M}$ represent a proper density matrix since it is normalized, Hermitian and positive-semidefinite. Note that if we are interested in some observables, the expectation value obtained from amplitude-matrix PT is not necessarily closer to the exact value than that from density-matrix PT. We will see in \cref{sec:examples} that whether the amplitude-matrix PT provides more accurate results depends on the particular perturbation.

The amplitude-matrix PT becomes more complicated if one or more eigenvalues of $\rsc{0}$ vanish, e.g.\ when $\rsc{0}$ represents a pure state. In that case, $\sZ$ in \cref{eq:zs_j} is non-invertible (see \cref{app:series_ze}) and thus a unique solution for \cref{eq:zs_j} does not exist (depending on the specific case, there could be infinitely many solutions or no solution). In the previous case in which there are infinitely many solutions, we may add an identity matrix component to $\rsc{0}$, i.e.\ $\rsc{0} \rightarrow \rsc{0} + c \openone$ where the parameter $c$ is small. The identity matrix acts as a correction matrix \cite{correction_matrix_1} which stabilizes the procedure of solving the linear equation [\cref{eq:zs_j}] to provide a unique $\zsc{j}$. An example of using the correction matrix is given in \cref{sec:single_qubit_resonator_ring}. If \cref{eq:zs_j} has no solution, other forms of series expansions would have to be applied. We will not further consider that case in the present paper. The correction matrix and the validity of the series expansion are discussed with more details in \cref{app:series_ze}.

\begin{figure}
\centering
\includegraphics[width=\columnwidth]{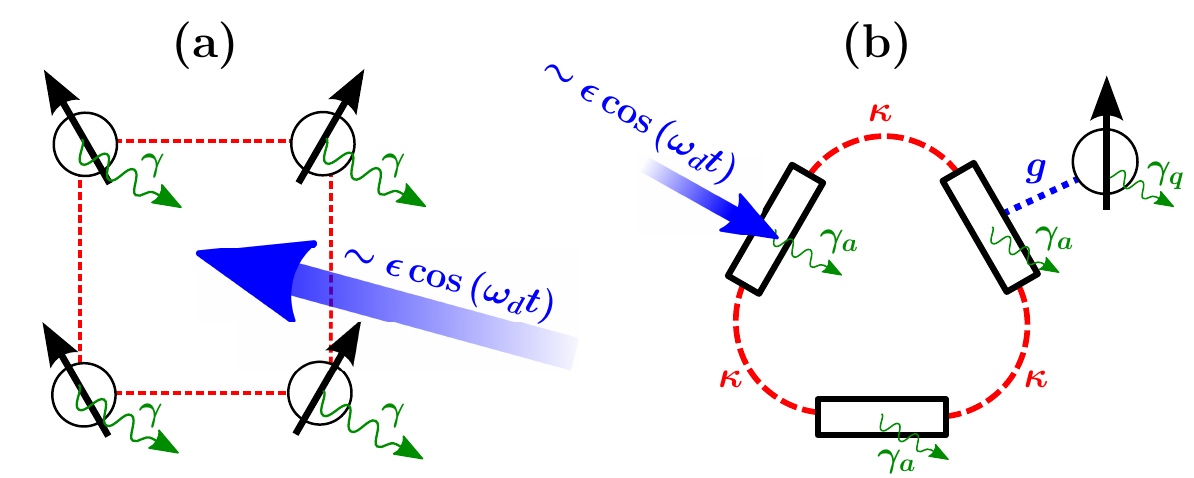}

	\caption{Schematic of the two open quantum systems discussed in \cref{sec:examples}. \textbf{(a)} The system consists of four spins coupled in a ring with nearest neighbor flip-flop coupling of strength $t$ (dashed red lines). We drive the spin ring coherently by a time dependent field (the big blue arrow). Moreover, each spin is subjected to local dissipation with spin relaxation rate $\gamma$ (green curly arrows). \textbf{(b)} The system is composed of a three-resonator ring (black rectangles) coupled to one qubit with an interaction strength $g$ (blue dotted line). The resonators are coupled to each other with a strength $\kappa$ (red dashed lines). One of the resonators is driven coherently (the blue arrow). Both the resonators and the qubit are subject to local dissipation with photon decay rate $\gamma_a$ and qubit relaxation rate $\gamma_q$ respectively (green curly arrows).}
	\label{fig:example}
\end{figure}

\begin{figure*}
\centering
\includegraphics[width=0.7\textwidth]{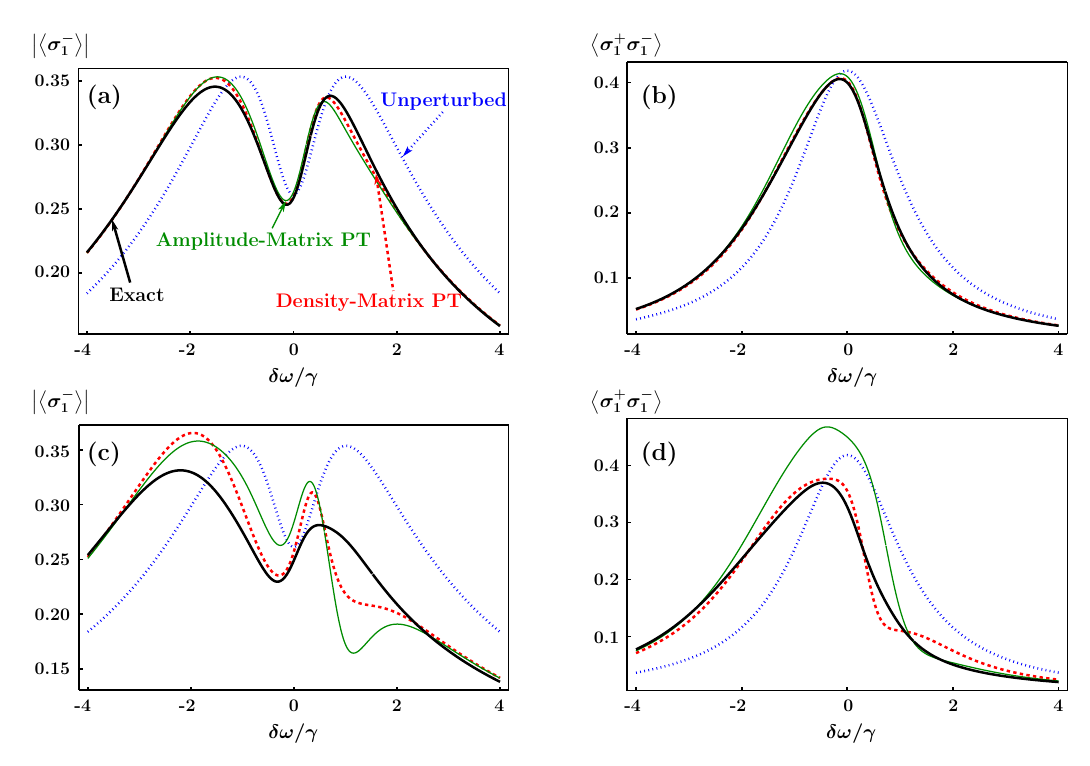}

	\caption{Shown are: the unperturbed result with respect to $\sL_0$ (blue dotted line), the exact result with respect to $\sL$ (black solid line), the second-order density-matrix-PT result (red dashed line) and the second-order amplitude-matrix-PT result (green thin line). \textbf{(a)} and \textbf{(b)}: $\abs{\avg{\s^-_1}}$ and $\avg{\s^+_1 \s^-_1}$ are plotted as a function of $\frac{\Dw}{\gamma}$ for $\frac{\epsilon}{\gamma}=0.8$ and $\frac{t}{\gamma}=0.4$. The exact result is well approximated by the second-order PT result. The amplitude-matrix PT is less accurate especially in $\avg{\s^+_1 \s^-_1}$. \textbf{(c)} and \textbf{(d)}: The same is plotted except for $\frac{t}{\gamma}=0.8$. Although the deviation is relatively large, the shape of the exact result is still qualitatively captured by the second-order PT results.}
\label{fig:a_spin_chain}
\end{figure*}

\section{Comparing PT with exact results}
\label{sec:examples}
In this section, we consider two examples (\cref{fig:example}) to illustrate the use and accuracy of PT. We compare steady-state expectation values from second-order PT with those from exact diagonalization of $\sL$. We then examine whether the expectation values from PT capture the shape of those from exact diagonalization. Note that the steady-state result obtained from density-matrix PT can be non-positive. This is indeed the case for some choices of parameters in the two following examples. Therefore, we also apply the amplitude-matrix PT and compare the results between two perturbative treatments. Here, although we choose small systems to enable exact diagonalization, the PT can be extended to a larger system without much difficulty.

\subsection{Four spins coupled in a ring}
\label{sec:1D_spin_chain}
We consider four spins coupled in a ring as shown in \cref{fig:example}(a). The spins are coupled by flip-flop interaction with spin-spin coupling strength $t$. We drive all the spins equally with a drive strength $\epsilon$ and a drive frequency $\omega_d$. Within the Rotating Wave Approximation (RWA), the Hamiltonian is given by
\begin{align}
H=&\sum_{n} \left[\Dw \, \s^+_n \s^-_n + \epsilon (\s^+_n+\s^-_n) \right]\nn\\
&+t \sum_{n} \left(\s^+_n \s^-_{n+1}+ \mathrm{h.c.}\right).
\label{eq:H_spin_chain}
\end{align}
Here, $\s^\pm_n$ is the raising or lowering operator of the spin at site $n$ and $\Dw \equiv \omega_0-\omega_d$ is the detuning between the energy splitting $\omega_0$ of the spin and the drive frequency $\omega_d$. Note that in \cref{eq:H_spin_chain}, the time dependence of the drive has already been eliminated by working in the rotating frame. All four spins are coupled to a zero-temperature bath. This leads to spin relaxation with relaxation rate $\gamma$. Thus, $\sL$ is given by
\be
\label{eq:L_spin_chain}
\sL \rho = -i [H,\rho] +\gamma \sum_{n} \sD\left[\s^-_n \right]\rho,
\ee
where $\sD\left[\s^-_n \right]\rho \equiv \s^-_n\rho \s^+_n-\frac{1}{2} \s^+_n \s^-_n \rho -\frac{1}{2} \rho \s^+_n \s^-_n $ is the usual dissipator for spin relaxation.

We now begin the perturbative treatment by separating $\sL$ into two parts, $\sL=\sL_0+t\, \sL_1$. Here, $\sL_0$ describes the ``atomic limit" in which the spin-spin coupling is absent, i.e.\ 
\begin{align*}
\sL_0 \rho
= \sum_{n}\{ -i \left[\Dw \s^+_n \s^-_n + \epsilon (\s^+_n+\s^-_n),\rho \right] +\gamma\,  \sD \left[\s^-_n \right]\rho \}.
\end{align*}
The perturbation $t \, \sL_1$ captures the spin-spin coupling,
\begin{align}
t\, \sL_1\rho= \, t\sum_{n} -i \left[\left( \s^+_n \s^-_{n+1}+\s^-_n \s^+_{n+1} \right),\rho \right],
\end{align}
and the order of perturbation is counted with respect to $t$. We next apply second-order PT to compute the steady-state expectation values for specific operators. We choose $\s^-_1$ and the excitation number $\s^+_1 \s^-_1$ because they represent the reduced density matrix of a single spin (which is the same for all spins due to symmetry). In \cref{fig:a_spin_chain}, we compare results from density-matrix PT and amplitude-matrix PT to the exact result and the unperturbed result.

We first consider the case shown in \cref{fig:a_spin_chain}(a) and (b) where the coupling strength $t$ represents the smallest energy scale. As discussed in Ref.\ \cite{circuitQED3}, we observe two symmetric resonance peaks of $\abs{\avg{\s^-_1}}$ in the unperturbed result. When the coupling is present, the two peaks are shifted in position and become asymmetric, in agreement with the results from Ref.\ \cite{coupled_JC2012}. For $\avg{\s^+_1 \s^-_1}$, we also observe the shift and the asymmetry of the resonance peak in the presence of coupling. The second-order PT well captures the above features. Note that the amplitude-matrix-PT result is slightly less accurate than the density-matrix-PT result.

To illustrate the limitation of PT, we next increase the coupling strength so that it equals the drive strength. We show result for this parameter choice in \cref{fig:a_spin_chain}(c) and (d). Qualitatively, the shape of the curves from the exact calculation is still captured by the perturbative results. However, the results from PT show relatively large deviations from the exact result. This is expected since the perturbation parameter $t$ is now roughly the same as both $\epsilon$ and $\gamma$, thus PT begins to break down.

\subsection{Single qubit coupled to a three-resonator ring}
\label{sec:single_qubit_resonator_ring}
As another example, we choose a system composed of a single qubit coupled to a three-resonator ring, see \cref{fig:example}(b). The resonators are coupled together with a photon hopping rate $\ka$. We drive one of the resonators with a drive strength $\epsilon$ and a drive frequency $\omega_d$. The qubit is coupled to another resonator with a coupling strength $g$. Within RWA, the system Hamiltonian $H$ is given by
\begin{align}
H=&\sum_n \left[ \Dw \, a\dg_n a_n + \ka \left( a_n\dg a_{n+1} +a_n a_{n+1}\dg \right) \right] + \epsilon \left(a_1 +a_1\dg \right) \nn\\
&+\Dw \, \s^+ \s^-
+g\left( a_2 \s^+ +a_2\dg \s^- \right).
\label{eq:H_scattering}
\end{align}
Here, $a_n$ ($a_n\dg$) is the annihilation (creation) operator of the resonator mode at site $n$ and $\Dw \equiv \omega-\omega_d$ is the detuning between the bare resonator and qubit frequency $\omega$ and the drive frequency $\omega_d$. Note that in \cref{eq:H_scattering}, the time dependence of the drive has been once again eliminated by working in the rotating frame. The qubit and resonators are each coupled to a zero-temperature bath. This leads to qubit relaxation and photon decay with rate $\gamma_q$ and $\gamma_a$, respectively. Thus, $\sL$ is given by
\be
\sL \rho=-i \left[H,\rho \right] +\gamma_a \sum_{n}\sD \left[a_n \right] \rho+\gamma_q \sD \left[\s^- \right] \rho.
\ee

The qubit-resonator coupling mediates two effects, an indirect coherent drive on the qubit and the correlation between resonator-ring and qubit subsystems. We treat the latter effect, i.e.\ the correlation, as the perturbation. Here, we separate the effect of coupling by making a coherent displacement as follows.
If the coupling between resonators and qubit is absent, the eigenmodes of the resonator ring are in the coherent state with $\avg{\at_\mu}^{(0)} \equiv \alpt_\mu $ where 
\be
\label{eq:amplitdue_decoupled}
\alpt_\mu=\frac{-\epsilon}{\sqrt{3}} \lf( \Dw+ 2\ka \cos \tfrac{2\pi\mu}{3} - i \tfrac{\gamma_a}{2} \rt)^{-1} e^{i \tfrac{2\pi \mu}{3}}.
\ee
Here, $\at_\mu \equiv \tfrac{1}{\sqrt{3}} \sum_{n} a_n \exp({i \tfrac{2\pi}{3} \mu n})$ is the annihilation operator of the eigenmodes and $\mu$ is the index labeling the eigenmodes.
We then displace $\at_\mu$ according to
\be
\label{eq:displacement}
\at'_\mu = \at_\mu - \alpt_\mu.
\ee
By using this displacement, we rewrite the Liouville superoperator as
\be
\sL \rho = -i \lf[H_0 + g H_1,\rho \rt] +\gamma_a \sum_{\mu}\sD \left[\at'_\mu \right] \rho+\gamma_q \sD \left[\s^- \right] \rho. \nn
\ee
Here, $H_0$ is the decoupled Hamiltonian (between the displaced eigenmodes $\at'_\mu$ and the qubit) given by
\begin{align}
H_0 =& \sum_\mu  \lf(\Dw+  2\ka \cos \tfrac{2\pi\mu}{3}\rt) \lf(\at'_\mu\rt)\dg \at'_\mu \nn\\
&+\Dw \, \s^+ \s^- + \lf(\epe \s^+ + \epe^* \s^- \rt)
\end{align}
where $\epe \equiv  \tfrac{g}{\sqrt{3}} \sum_{\mu} \alpt_\mu \exp({-i \tfrac{4 \pi \mu}{3}})$ is the effective drive on the qubit. The remaining part $g H_1$ describing the coupling between the displaced eigenmodes and the qubit is given by
\be
g H_1=\frac{g}{\sqrt{3}} \sum_{\mu}\left(e^{-i \tfrac{4 \pi \mu}{3}} \at'_\mu \s^+  + \mathrm{h.c.} \right).
\ee

\begin{figure*}
\centering
\includegraphics[width=\textwidth]{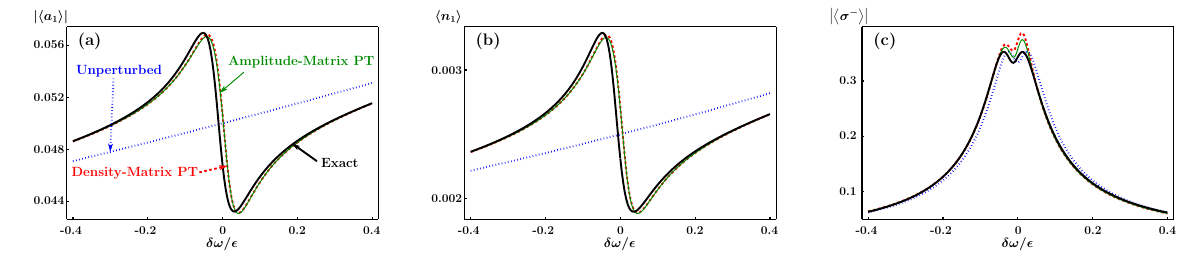}

	\caption{The color scheme of \cref{fig:a_spin_chain} is used. To apply the amplitude-matrix PT, the correction matrix is employed with the parameter $c=10^{-9}$. We plot $\abs{\avg{a_1}}$, $\avg{n_1}$ and $\abs{\avg{\s^-}}$ as a function of $\frac{\Dw}{\epsilon}$ for $\frac{\ka}{\epsilon}=10$, $\frac{g}{\epsilon}=0.5$ and $\frac{\gamma_q}{\epsilon}=\frac{\gamma_a}{\epsilon}=0.05$. The resonance at $\Dw=0$ is well captured by second-order PT. The amplitude-matrix PT and density-matrix PT perform with nearly identical accuracy.}
\label{fig:a_scattering}
\end{figure*}

Now, we begin the perturbative treatment by separating $\sL$  into two parts, $\sL=\sL_0 + g\, \sL_1$. The unperturbed superoperator $\sL_0$ is given by
\be
\sL_0 \rho=-i \left[H_0,\rho \right] +\gamma_a \sum_{\mu}\sD \left[\at'_\mu \right] \rho+\gamma_q \sD \left[\s^- \right] \rho.
\ee
The perturbation $g\, \sL_1$ is given by
\be
g\, \sL_1 \rho = -i \left[g H_1,\rho \right],
\ee
and the order of perturbation is counted with respect to $g$.
The steady state of $\sL_0$ is a product state composed of density matrices of the resonator ring and the qubit respectively. The resonator ring is in a pure state with the displaced eigenmodes in the vacuum state. As a consequence, the unperturbed density matrix $\rsc{0}$ has multiple eigenvalues zero. Therefore, to apply the amplitude-matrix PT, we employ the correction matrix discussed in \cref{sec:amplitude_matrix} (with parameter $c=10^{-9}$).

We are interested in the expectation values at site 1, specifically $\avg{a_1}$ and $\avg{n_1} \equiv \savg{a_1\dg a_1}$, as a function of the drive frequency expressed in terms of $\Dw$. For the system in which the resonator-ring and the qubit are decoupled ($g=0$), we expect two resonances at $\Dw=-2\ka$ and $\Dw=\ka$ corresponding to the eigenmodes of the resonator ring. Once the qubit is coupled to the resonator ring, we expect a resonance at $\Dw=0$ originating from the qubit's response to the drive. We monitor this response by calculating the expectation value of $\s^-$

We can now verify that the resonance at $\Dw=0$, a key consequence of the coupling, is successfully captured by second-order PT. To this end, we consider the case that the drive and coupling strengths $\epsilon$ and $g$ are small compared to the hopping rate $\ka$ but large compared to the relaxation and decay rates $\gamma_q$ and $\gamma_a$. Note that the perturbation parameter $g$ is \textit{not} the smallest energy scale in this case. Nonetheless, we will see that PT still holds. The expectation values of $a_1$, $n_1$ and $\s^-$ are shown in \cref{fig:a_scattering}(a), (b) and (c) respectively. The results from second-order PT match well with the exact result and the amplitude-matrix PT and density-matrix PT give results with nearly identical accuracy. The saturation effect visible in \cref{fig:a_scattering}(c) shows that the qubit is in the nonlinear regime. Note that the correction matrix method is reliable. This is demonstrated by the consistency between the amplitude-matrix-PT result and the exact result.

\section{Conclusion}
\label{sec:conclusion}
We investigated the perturbative approach to Markovian open quantum systems and developed a non-degenerate PT based on the quantum master equation. The density-matrix PT recursively determines the corrections to the eigenvalues and right eigenstates of the Liouville superoperator $\sL$. This perturbative scheme may lead to non-positive steady-state ``density matrices" as a result of truncation. This makes a key difference between density-matrix PT and closed-system PT, which always yields proper quantum states. The issue of non-positivity can be tackled by a modified perturbative scheme based on the amplitude matrix. With two example systems, we illustrated that the approximate results are in excellent agreement with exact results for representative parameter choices. The expectation values obtained from density-matrix PT showed good agreement in the two examples even if the truncated density matrix was slightly non-positive.

The perturbative treatment presented here is suitable for systems of sizes that cannot be handled by exact solution of the quantum master equation. An interesting future application of this PT consists of the study of open quantum systems with a lattice structure, such as the open Jaynes-Cummings lattice. Promising experimental progress \cite{circuitQED4} indicates that such open lattices can indeed be implemented in the circuit QED architecture, and serve as open-system quantum simulators \cite{quantum_simulator_1}. Openness and relatively large size of such systems make the theoretical investigation challenging. We believe that the developed open-system PT will provide a useful tool in studying the physics of open lattice systems.

\begin{acknowledgments}
We thank Guanyu Zhu and Joshua Dempster for valuable discussions. This research was supported by the NSF under Grant No. PHY-1055993 (A.C.Y.L. and J.K.) and by the South African Research Chair Initiative of the Department of Science and Technology and National Research Foundation (F.P.). J.K. and F.P. thank the National Institute for Theoretical Physics for supporting J.K.’s visit through a NITheP visitor program grant.
\end{acknowledgments}

\appendix
\section{Review of the Moore-Penrose pseudoinverse}
\label{app:pseudoinverse}
In this Appendix, we review the definition and basic properties of the Moore-Penrose (MP) pseudoinverse following Refs.\ \cite{campbell1991,pseudoinverse2012}. The notion of the MP pseudoinverse was introduced by E.~H.~Moore in 1920 \cite{Moore1920} and then independently by R.~Penrose in 1955 \cite{Penrose1955} to deal with matrices that have no inverse in the ordinary sense.

Before we review the formal definition, we motivate the MP pseudoinverse by the following simple consideration. The inverse $\nInverse{A}$ is well-defined if and only if $A$ is a non-singular square matrix. However, even if the matrix is singular, we sometimes need to find a generalized inverse which resembles the normal inverse, for example in the ordinary perturbation theory. If a square matrix $A$ is Hermitian, the natural way to do this is as follow. First, factorize $A$ in the form
\be
A = U D \, U\dg,
\ee
where $U$ is unitary and $D$ is a diagonal matrix which consists of the eigenvalues of $A$. Then, a matrix $D'$ is defined by taking the reciprocal of each non-zero diagonal element while leaving the zero elements unchanged. A generalized inverse $A'$ can thus be defined as
\be
\label{eq:Hermitian_generalized_inverse1}
A' = U D' U\dg.
\ee

This becomes more complicated when the square matrix $A$ is not Hermitian, an example being the superoperator $\sL_0$. For our purpose, we assume that $A$ is diagonalizable such that
\be
\label{eq:non_Hermitian_decomposition}
A= Y D \nInverse{Y},
\ee
where $Y$ is non-unitary. It is tempting to define the generalized inverse $A'$ in the form similar to \cref{eq:Hermitian_generalized_inverse1} such that
\be
\label{eq:generalized_inverse1}
A'= Y D' \nInverse{Y}.
\ee
However, this definition of generalized inverse is not closest to the normal inverse. Recall that if $A$ is invertible, $A \nInverse{A} = \openone = \nInverse{A} A$ means that $A \nInverse{A} $ and $\nInverse{A} A$ are Hermitian. By directly using \cref{eq:generalized_inverse1}, we show that neither $A' A$ nor $A A'$ is Hermitian.

To achieve this, recall that the singular value decomposition (SVD) provides an alternative to \cref{eq:non_Hermitian_decomposition} of relating a non-Hermitian matrix $A$ to a diagonal matrix $D$, namely
\be
\label{eqb:SVD}
A = U D V\dg,
\ee
where $U$ and $V$ are unitary matrices. Note that the diagonal matrices in \cref{eq:non_Hermitian_decomposition,eqb:SVD} are not the same. Similar to the treatment for Hermitian singular matrices, the pseudoinverse (or the generalized inverse) $A\pinv$ for a non-Hermitian matrix $A$ is defined as
\be
\label{eq:pinvSVD}
A\pinv = V D\pinv U\dg.
\ee
Here, $D\pinv$ is defined by taking the inverse of the non-zero diagonal elements of $D$.

There are three advantages to use $A\pinv$ in \cref{eq:pinvSVD} instead of $A'$ in \cref{eq:generalized_inverse1}. First of all, $A\pinv A$ and $A A\pinv$ are now projectors that are Hermitian. Secondly, it is computationally efficient to calculate the SVD and thus also the pseudoinverse $A\pinv$. Finally, due to the fact that the SVD is defined for any complex-valued matrix, we can easily generalize the pseudoinverse $A\pinv$ to cases where $A$ cannot be diagonalized or is not a square matrix. In fact, the pseudoinverse defined in \cref{eq:pinvSVD} is called the MP pseudoinverse.

Now, we prove one property of the MP pseudoinverse which is used in \cref{app:proofperturbation}. The claim is 
\be
\label{eq:pinv5}
A A\pinv = P_{A},
\ee
where $P_{A}$ is the orthogonal projector onto the range of $A$. By using the definition of $A\pinv$ in \cref{eq:pinvSVD}, we obtain
\be
\label{eqb:defintion_1}
A A\pinv  A =A.
\ee
This implies $(A A\pinv)^2 = A A\pinv$, i.e.\ $A A\pinv$ is a projector. Likewise, we infer from \cref{eq:pinvSVD} that
\be
A A\pinv = (A A\pinv)\dg
\ee
which means $A A\pinv$ is Hermitian. Since a projector is Hermitian if and only if it is an orthogonal projector, $A A\pinv$ is an orthogonal projector. Let us denote the range of $A$ by $\Range{A}$. Then, \cref{eqb:defintion_1} immediately yields $\Range{A} = \Range{A A\pinv A}$. Using the fact that composite maps decrease the range according to $\Range{AB} \subseteq \Range{A}$, we infer that
\begin{align}
\Range{A} \subseteq \Range{A A\pinv} \subseteq \Range{A}
\end{align}
The above relation can only hold if and only if $\Range{A}= \Range{A A\pinv}$. Thus we proved $A A\pinv$ is an orthogonal projector with the same range as $A$, i.e.\ $A A\pinv = P_{A}$.

\section{Proof of the general form of perturbative corrections}
\label{app:proofperturbation}
In this Appendix, we provide details of the derivation leading to \cref{eq:evrj,eq:ev0j,eq:rsj} in the main text.

We first wish to prove that the expression of $\evrmc{j}$ in \cref{eq:evrj}, rewritten as
\begin{align}
\label{eqa:evrj}
\evrmc{j} &= \left(  \sL_0 -\evamc{0} \right)\pinv \rem,
\end{align}
is indeed a solution to \cref{eq:linear_algebra_recursive} rewritten in the form
\be
\label{eqa:linear_algebra_recursive}
\left(\sL_0 - \evamc{0} \right) \evrmc{j}  = \rem,
\ee
where $\rem = -\sL_1 \evrmc{j-1} +\Demc{j}$.
Solving \cref{eqa:linear_algebra_recursive} for $\evrmc{j}$ is a standard linear algebra problem. The necessity for working with a pseudoinverse lies in the fact that $( \sL_0 -\evamc{0})$ is singular and non-Hermitian. This prevents us from using the normal inverse to solve for $\evrmc{j}$. Here, we employ the Moore-Penrose pseudoinverse denoted by $\pinv$ (see \cref{app:pseudoinverse}). This choice is useful because calculating the Moore-Penrose pseudoinverse is computationally efficient by means of the singular value decomposition.

After plugging $\evrmc{j}$ from \cref{eqa:evrj} into \cref{eqa:linear_algebra_recursive}, it is clear that the proof amounts to verifying that
\begin{align}
 \lf(\sL_0 - \evamc{0} \rt)  \lf(\sL_0- \evamc{0}\rt)\pinv \rem =\rem.
\label{eq:Aproofeq0}
\end{align}
Note that if  $(\sL_0 - \evamc{0})$ was invertible, the proof would be trivial. Here, we will need to rely on the special properties of the Moore-Penrose pseudoinverse. With the help of \eqref{eq:pinv5}, we rewrite \cref{eq:Aproofeq0} in the equivalent form,
\begin{align}
\sP_{\sL_0- \evamc{0}} \, \rem = \rem.
\label{eq:Aproofeq2}
\end{align}
where $\sP_{\sL_0- \evamc{0}}$ is the orthogonal projector onto the range of $(\sL_0- \evamc{0})$. Recall that any projector acts as the identity matrix on vectors from its range, which means that \cref{eq:Aproofeq2} holds if $\rem$ belongs to the range $\Ran$ of $\sP_{\sL_0- \evamc{0}}$. In addition, note that since $\sP_{\sL_0- \evamc{0}}$ is an orthogonal projector, the range $\Ran$ and the null space $\Ker$ of $\sP_{\sL_0- \evamc{0}}$ are orthogonal spaces, i.e.\ $\sinner{x}{y}=0$ $\forall x\in \Ran$ and $\forall y\in \Ker$. This implies that $\rem$ belongs to $\Ran$ if and only if $\rem$ is orthogonal to $\Ker$. To conclude the proof, we therefore only need to show that $\rem$ is orthogonal to the null space $\Ker$ of $\sP_{\sL_0- \evamc{0}}$.

To prove the above statement, we claim a lemma that $\Ker$ is spanned by the left eigenstate $\evlmc{0}$ of $\sL_0$ with eigenvalue $\evamc{0}$, i.e.\ 
\be
\label{eqa:lemma}
\Ker = \textrm{span}\{\evlmc{0} \}.
\ee
We will prove this lemma later in this paragraph. Note that the recursive relation $\eqref{eq:evaj}$ can be rewritten in the form
\be
\label{eq:Aproofeq1}
\inner{\evlmc{0}}{\rem}=0,
\ee
which means $\rem$ is orthogonal to $\evlmc{0}$. Together with lemma \eqref{eqa:lemma}, this proves that $\rem$ is orthogonal to the null space $\Ker$ of $\sP_{\sL_0- \evamc{0}}$.
Now, to complete the argument, we prove lemma \eqref{eqa:lemma} as follows. Since $\evlmc{0}$ is the left eigenstate of $\sL_0$ with eigenvalue $\evamc{0}$, it is clear that $\sinner{\evlmc{0}}{(\sL_0- \evamc{0}) A}=0$ for any matrix $A$. This means $\evlmc{0}$ is orthogonal to the range of $(\sL_0- \evamc{0})$. Since $(\sL_0- \evamc{0})$ and $\sP_{\sL_0- \evamc{0}}$ share the same range, $\evlmc{0}$ is also orthogonal to the range $\Ran$ of $\sP_{\sL_0- \evamc{0}}$.  Thus, $\evlmc{0}$ is an element of the null space $\Ker$ of $\sP_{\sL_0- \evamc{0}}$. Assuming that $\evamc{0}$ is a non-degenerate eigenvalue of $\sL_0$, we can further infer that $\Ker$ is spanned by $\evlmc{0}$, which is lemma \eqref{eqa:lemma}.
Therefore, we proved that the form of $\evrmc{j}$ in \cref{eqa:evrj} serves as a solution of \cref{eqa:linear_algebra_recursive}, i.e.\ \cref{eq:evrj} is the recursive relation for corrections to right eigenstates.

In particular, if we are interested in the steady-state density matrix $\rs$ defined by $\sL \rs =0$, i.e.\  $\rs \equiv \evr{0}$, we can simplify the recursive relations as following. Let us recall that $\sL$ and $\sL_0$ are proper generators of quantum dynamical semigroups. In order to be trace preserving (which is a necessary condition of a proper generator), the identity $\openone$ is the left eigenstate of $\sL$ and $\sL_0$ with eigenvalue zero, i.e.\ $\evl{0}=\evlc{0}{0}=\openone$. It follows that $\openone$ is also the left eigenstate of $\sL_1$ with eigenvalue zero since $\sL_1=\sL-\sL_0$. And thus $\Trb{\openone \sL_1 x}=0$ for any operator $x$. It is straightforward to show from $\Trb{\openone \sL_1 x}=0$ and \cref{eq:evaj,eq:evrj} that $\forall j \in \mathbb{N}$,
\begin{align}
\label{eq:Aev0j}
\evac{0}{j} =& 0, \\
\label{eq:Arsj}
\rsc{j}=& -\sL_0\pinv  \sL_1 \rsc{j-1}.
\end{align}

\section{Validity of the series expansion of amplitude matrices}
\label{app:series_ze}
The series expansion of $\zs$ is valid if all $\zsc{j}$ can be determined according to 
\be
\sZ \zsc{j} = \rsc{j}- \sum_{k=1}^{j-1} \zsc{k}  (\zsc{j-k})\dg,
\label{eqc:zs_j}
\ee
which is \cref{eq:zs_j} in the main text. Note that $\sZ$ is defined by $\sZ \bullet = \zsc{0} (\bullet)\dg + \bullet (\zsc{0})\dg $ and $\zsc{0}$ is determined through Cholesky decomposition: $\rsc{0} = \zsc{0} (\zsc{0})\dg $. \Cref{eqc:zs_j} is a system of linear equations and thus it has a unique solution if and only if $\sZ$ is invertible.

Whether $\sZ$ is invertible thus depends on the form of $\zsc{0}$ (which ultimately depends on $\rsc{0}$ which is Hermitian and positive-semidefinite). If one of the eigenvalues of $\rsc{0}$ is zero, there is a corresponding eigenvector $\ket{\psi}$ such that $\rsc{0}\ket{\psi}=0$. Consider the decomposition: $\rsc{0} = h h$ where $h$ is a Hermitian matrix \footnote{The existence of the decomposition, $\rsc{0} = h h$, can be proven by writing $\rsc{0}$ in its eigen-decomposition form: $\rsc{0} = U D U\dg$ where $U$ is a unitary matrix and $D$ is a diagonal matrix with non-negative elements. Thus, $h$ can be defined as $h = U \sqrt{D} U\dg$ where $\sqrt{D}$ is defined as taking the square root of each diagonal element of $D$.}. Since $\ket{\psi}$ is the eigenvector of $\rsc{0}$ with eigenvalue zero, $\ket{\psi}$ is also the eigenvector of $h$ with eigenvalue zero, i.e.\ $h\ket{\psi}=0$. Moreover, due to the fact that $\zsc{0} (\zsc{0})\dg = \rsc{0} = h h$, $\zsc{0}$ and $h$ are unitarily right equivalent \cite{matrix_mathematics_2009}, i.e.\ $\zsc{0} = h S$ where $S$ is a unitary matrix. Thus, $\ket{\psi}$ is also the left eigenvector of $\zsc{0}$ with eigenvalue zero, i.e.\ $(\zsc{0})\dg \ket{\psi} = S\dg h\ket{\psi}  =0$. Now, there must be a right eigenvector of $\zsc{0}$, denoted by $\ket{\phi}$, that corresponds to the same eigenvalue (which is zero), i.e.\ $\zsc{0} \ket{\phi} =0$. We can show by directly substitution that $\sZ(T \ket{\phi}\bra{\phi} )=0$ where $T$ is the matrix corresponding to Gaussian elimination which transforms $ (\ket{\phi}\bra{\phi} )$ to a lower triangular matrix. Therefore, $\sZ$ is not invertible if $\rsc{0}$ contains at least one eigenvalue zero.

If $\sZ$ is not invertible, \cref{eqc:zs_j} have infinitely many solutions or no solution. The former case, in which there are infinitely many solutions, can be bypassed if we can make the eigenvalues of $\rsc{0}$ non-zero. In order to avoid the eigenvalue zero, we consider to shift $\rsc{0}$ by an identity matrix component according to
\be
\rsc{0} \rightarrow \rsc{0} + c \openone.
\label{eqc:corr_matrix}
\ee
Here, we choose the parameter $c$ such that the result is stable with respect to variation of $c$ and we expect that $c$ tends to zero. In this way, the procedure in solving \cref{eqc:zs_j} is stabilized and we obtain a unique $\zsc{j}$. In fact, the method described above is similar to the correction matrix method \cite{correction_matrix_1,correction_matrix_2} for the Cholesky decomposition of matrices with eigenvalue(s) zero. There, a small diagonal correction matrix is also added to the original matrix to avoid the eigenvalue(s) zero. And thus, $c \openone$ in \cref{eqc:corr_matrix} corresponds to the correction matrix.

If we indeed encounter the latter case in which there is no solution, it means that $\zsc{j}$ cannot be determined. In fact, a two-level system coupled to finite temperature bath with $\sD \lf[\s^+ \rt]$ term treated as the perturbation belongs to this case. The failure of determining all $\zsc{j}$ means that the series expansion of $\zs$ is invalid. This originates from the fact that if $\rsc{0}$ contains any zero eigenvalues, the leading order term of some elements of $\zs$ may be of the order of $\alpha^{1/2}$ (or $\alpha^{3/2}$, etc.) instead of $\alpha^0$. For real functions, an analogy would be the case $y(\alpha)=x^2(\alpha)$ where $y$ can be written as a power series in $\alpha$. If the leading order term of $y$ is of the order of $\alpha$, $x$ is a series that only contains half-integer orders of $\alpha$ and thus it is not a power series anymore.  This suggests us to try expansions in other form. We will not further discuss this case in the present paper.

\bibliography{cit}
\end{document}